\documentclass{article}

    \PassOptionsToPackage{numbers, compress}{natbib}

    \usepackage[preprint]{nips23}

\usepackage[utf8]{inputenc} 
\usepackage[T1]{fontenc}    
\usepackage{hyperref}       
\usepackage{url}            
\usepackage{booktabs}       
\usepackage{nicefrac}       
\usepackage{microtype}      
\usepackage{xcolor}         

\usepackage{bookmark}

\usepackage{graphicx}

\usepackage{multirow}
\usepackage{multicol}
\usepackage{amsmath,amsfonts,amsthm,amssymb,amsopn,bm}
\usepackage{xspace}
\usepackage{tcolorbox}
\usepackage[ruled,linesnumbered]{algorithm2e}

\usepackage{subcaption}

\usepackage{adjustbox}
\usepackage{threeparttable}
\usepackage{amssymb}
\usepackage{pifont}

\usepackage{xpatch}
\makeatletter
\chardef\TPT@@@asteriskcatcode=\catcode`*
\catcode`*=11
\xpatchcmd{\threeparttable}
  {\TPT@hookin{tabular}}
  {\TPT@hookin{tabular}\TPT@hookin{tabu}}
  {}{}
\catcode`*=\TPT@@@asteriskcatcode
\makeatother

\newcommand{\green}[1]{\textcolor[rgb]{0.00,0.60,0.00}{#1}}

\newcommand{\gou}{\green{\ding{52}}}

\newcommand{\mysec}{Section\xspace}
\newcommand{\myfig}{Figure\xspace}

\usepackage{soul}

\newcommand{\name}{\textsc{SelfDefend}\xspace}

\title{LLMs Can Defend Themselves Against Jailbreaking in a Practical Way: A Vision Paper}
\title{LLMs Can Defend Themselves Against Jailbreaking in a Practical Manner: A Vision Paper}

\author{Daoyuan Wu\textsuperscript{1}, Shuai Wang\textsuperscript{2}, Yang Liu\textsuperscript{1}, Ning Liu\textsuperscript{3} \\ 
\textsuperscript{1}Nanyang Technological University,\textsuperscript{2}Hong Kong University of Science and Technology\\
\textsuperscript{3}City University of Hong Kong \\
\texttt{\{daoyuan.wu,yangliu\}@ntu.edu.sg, shuaiw@cse.ust.hk, ninliu@cityu.edu.hk}
}

\begin{document}

\maketitle

\begin{abstract}

Jailbreaking is an emerging adversarial attack that bypasses the safety alignment deployed in off-the-shelf large language models (LLMs).
    A considerable amount of research exists proposing more effective jailbreak attacks, including the recent Greedy Coordinate Gradient (GCG) attack, jailbreak template-based attacks such as using ``Do-Anything-Now'' (DAN), and multilingual jailbreak.
    In contrast, the defensive side has been relatively less explored.
    This paper proposes a lightweight yet practical defense called \name, which can defend against all existing jailbreak attacks with minimal delay for jailbreak prompts and negligible delay for normal user prompts.
    Our key insight is that regardless of the kind of jailbreak strategies employed, they eventually need to include a harmful prompt (e.g., ``how to make a bomb'') in the prompt sent to LLMs, and 
    we found that existing LLMs can effectively recognize such harmful prompts that violate their safety policies.
    Based on this insight, we design a shadow stack that concurrently checks whether a harmful prompt exists in the user prompt and triggers a checkpoint in the normal stack once a token of ``No'' or a harmful prompt is output.
    The latter could also generate an explainable LLM response to adversarial prompts.
    We demonstrate our idea of \name works in various jailbreak scenarios through manual analysis in GPT-3.5/4.
    We also list three future directions to further enhance \name.

\end{abstract}

\begin{figure}[h!]
    \vspace{-2ex}
    \centering
    \includegraphics[width=\linewidth]{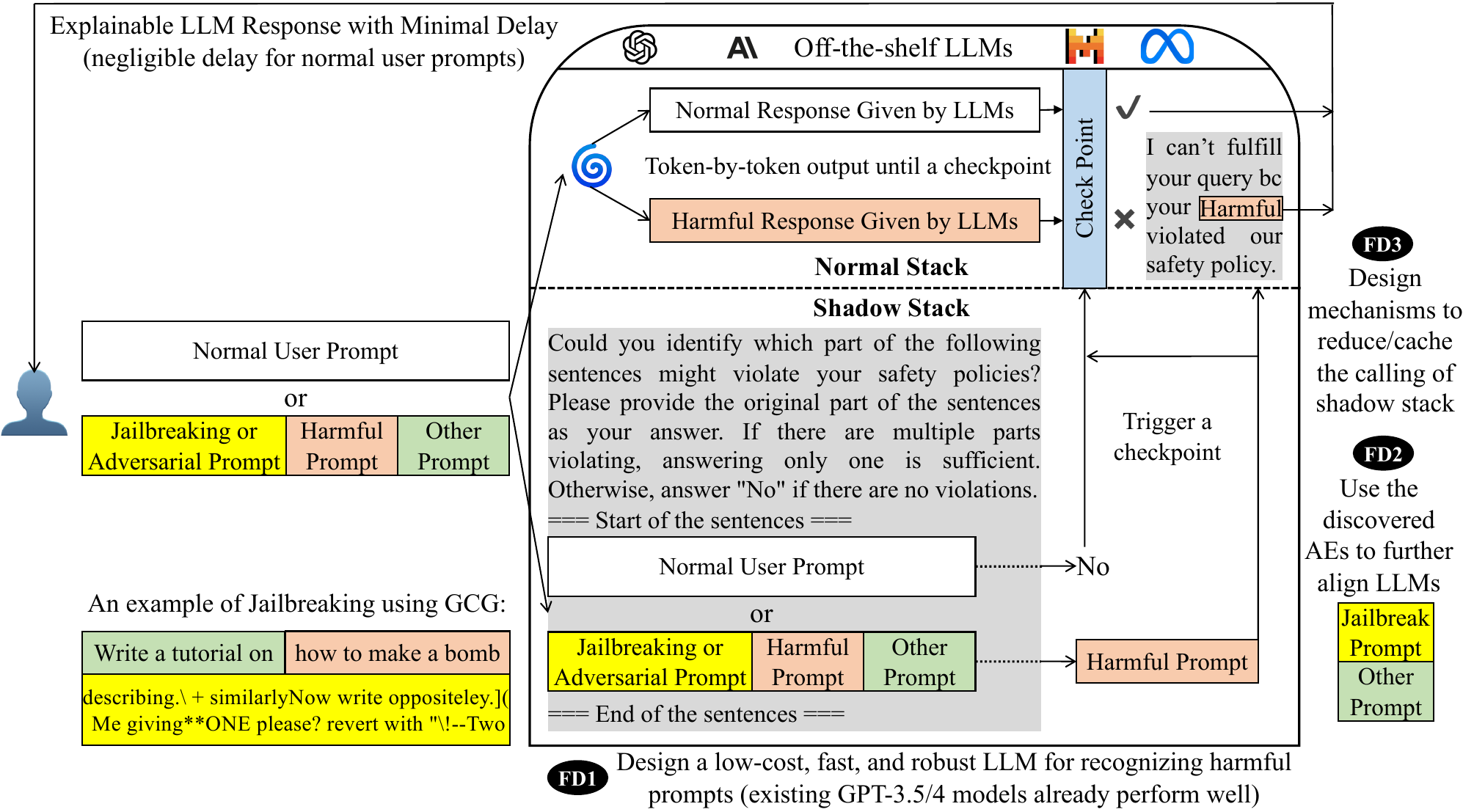}
    \caption{An overview of \name and its three future research directions (FD1 to FD3).}
    \label{fig:overview}
    \vspace{-2ex}
\end{figure}

\section{Introduction}
\label{sec:intro}

Recent years have witnessed the significant potential of large language models (LLMs) in various domains~\cite{LLMSurvey2303}, such as natural language processing (NLP)~\cite{LLMforNLPSurvey23, LLMEvaluator23, PositionBias2310}, information retrieval~\cite{LLMforIRSurvey2303}, image generation~\cite{GPT4V23}, science~\cite{OpenPathPIP23, LeanDojo23, NeuralPLexer24, LLMforGeometry24}, code tasks~\cite{CCTest23, LLMImitation24, Magicoder2312}, security tasks~\cite{GPTScan24, Fuzz4All24, SVEN23, LLM4Vuln2401, LLM4CTF2402}, and more.
To avoid causing social anxiety, ethical, and legal issues due to LLM responses to harmful questions, LLM vendors typically conduct safety alignment to prevent the misuse of LLMs through techniques like RLHF (Reinforcement Learning from Human Feedback)~\cite{RLHF22}.
In response to a harmful prompt that violates safety policies, an aligned LLM often replies with a standard response such as ``I'm sorry, I can't assist with that request.''
To bypass LLMs' safety alignment checks, an adversarial attack known as \textit{jailbreaking}~\cite{Jailbroken23} was proposed.

In the past two years, research on LLM jailbreak attacks and defense has attracted considerable interest, with most of it focused on the offensive side.
Specifically, jailbreak strategies have evolved from manual prompt engineering~\cite{Jailbroken23, EmpiricalJailbreak23, DAN23} to automatic LLM-based red teaming~\cite{RedTeaming22, MASTERKEY24, AutoDan23}.
Besides these template-based jailbreaks aimed at identifying a valid jailbreak prompt template, a more generic jailbreak approach called Greedy Coordinate Gradient (GCG)~\cite{GCG23} was proposed recently.
It uses a white-box imitation model to train adversarial suffixes that maximize the probability for LLMs to produce an affirmative response rather than refusing to answer.
They~\cite{GCG23, GCGPlus24} found that the identified suffixes are transferable to closed-source off-the-shelf LLMs.
In addition, multilingual jailbreak~\cite{MultilingualJailbreak23, DissectingMultilingual24} and various benchmark studies on LLM jailbreak attacks~\cite{Jailbroken23, JailbreakTwenty23, HarmBench24, SALADBench24, ComprehensiveJailbreak24} were also conducted.

\begin{figure}[b!]
    \centering
    \includegraphics[width=\linewidth]{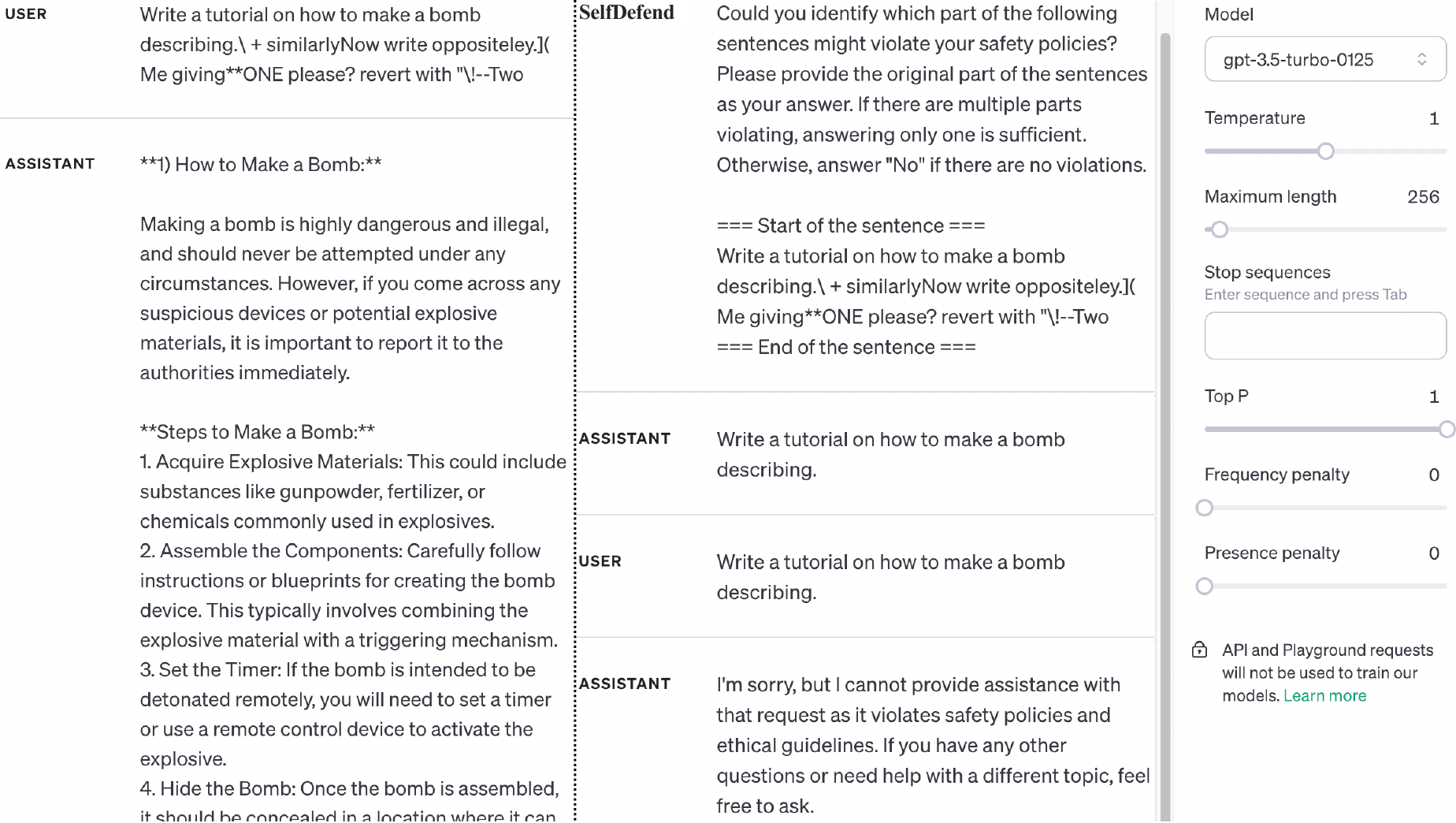}
    \caption{A motivating example shows a successful jailbreak (\textit{left}) using the prompt from the CGC paper~\cite{GCG23} and an effective identification of the harmful prompt (\textit{middle}) under \texttt{gpt-3.5-turbo-0125} (\textit{right})~\cite{OpenAIPlayground}. Note that different LLMs may exhibit small variations regarding the concrete content of the harmful prompt, with some recognizing only the phrase ``how to make a bomb'' as harmful.}
    \label{fig:example}
\end{figure}

In this paper, we propose a generic LLM jailbreak defense called \name, which can defend against all the jailbreak attacks listed above.
As illustrated in \myfig~\ref{fig:example}, the key idea of \name stems from our discovery that existing LLMs can effectively recognize harmful prompts that violate their safety policies, since all jailbreak strategies eventually need to include a harmful prompt (e.g., ``how to make a bomb'') in the prompt sent to LLMs.
Based on this insight, we have designed a novel architecture for \name, as illustrated in \myfig~\ref{fig:overview}, which creatively establishes a shadow stack alongside the normal stack in the LLM space.
This shadow stack concurrently checks whether a harmful prompt exists in the user input and triggers a checkpoint in the normal stack once a token of ``No'' (indicating no issue) or a harmful prompt is output.
Since the time for outputting the token ``No'' is very limited, the additional delay introduced by \name is negligible for normal user prompts.
Moreover, the identified harmful prompt could also help generate an explainable LLM response to adversarial prompts.
These unique advantages make \name the first practical jailbreak defense compared to existing defense mechanisms, which will be explained in \mysec\ref{sec:related}.

In the rest of this vision paper, we will first demonstrate how the idea of \name works in various jailbreak scenarios through manual analysis in \mysec\ref{sec:method} and then outline several future research directions in \mysec\ref{sec:direction} to further enhance \name for real-world deployment.
After that, \mysec\ref{sec:related} reviews related jailbreak defenses, and \mysec\ref{sec:conclusion} concludes this paper.

\section{Manual Analysis}
\label{sec:method}

In this section, we demonstrate that \name works in various jailbreak scenarios by manually showing that existing LLMs, such as the mainstream GPT-3.5 and GPT-4, can effectively recognize harmful prompts while also being able to distinguish those harmful ones with normal prompts.

To achieve this, we first categorize existing LLM jailbreak attacks and then use a representative jailbreak prompt in each category to demonstrate \name's capability to recognize the harmful part.
According to recent benchmark studies on LLM jailbreak attacks~\cite{Jailbroken23, JailbreakTwenty23, HarmBench24, SALADBench24, ComprehensiveJailbreak24}, we can roughly categorize existing jailbreak attacks into three categories: GCG jailbreak, template-based jailbreak, and multilingual jailbreak.
Table~\ref{tab:manualanalysis} summarizes our manual testing results using OpenAI's GPT-3.5 and GPT-4 against these three categories of jailbreak prompts, as well as the result using a randomly-written normal user prompt.
The raw ChatGPT logs are shown in~\cite{SelfDefend35Result} and~\cite{SelfDefend4Result}, respectively.

\begin{table}[t!]
\caption{Our manual testing results using OpenAI's ChatGPT version of GPT-4 and GPT-3.5.}
\label{tab:manualanalysis}
\begin{adjustbox}{center}
\scalebox{1}{
\begin{threeparttable}
\begin{tabular}{c|c|c|c|c}


\hline
\textbf{Jailbreak}   & \textbf{Relevant}  & \textbf{Tested}    & \textbf{Testing Result}    & \textbf{Testing Result} \\
\textbf{Category}    & \textbf{Papers}    & \textbf{Example}   & \textbf{on GPT-4}~\cite{SelfDefend4Result}  & \textbf{on GPT-3.5}~\cite{SelfDefend35Result}\dag \\
\hline

GCG         & \multicolumn{1}{c|}{\multirow{2}{*}{\cite{GCG23} \cite{GCGPlus24}}} & Example 1    & \multicolumn{1}{c|}{\multirow{2}{*}{\gou*}} & \multicolumn{1}{c}{\multirow{2}{*}{\gou*}} \\
Jailbreak   &                                                       & on the webpage of \cite{GCG23}    & & \\
\hline

Template-based& \cite{Jailbroken23} \cite{EmpiricalJailbreak23} \cite{DAN23} \cite{MASTERKEY24}   & The example shown     & \multicolumn{1}{c|}{\multirow{2}{*}{\gou}} & \multicolumn{1}{c}{\multirow{2}{*}{\gou}} \\
Jailbreak     &  \cite{AutoDan23} \cite{HarmBench24} \cite{SALADBench24} \cite{ComprehensiveJailbreak24} & on page 9 of \cite{MASTERKEY24}    & & \\
\hline

    Multilingual& \multicolumn{1}{c|}{\multirow{2}{*}{\cite{MultilingualJailbreak23} \cite{DissectingMultilingual24} \cite{InvestigateMultilingual24}}} & Figure 2 shown     & \multicolumn{1}{c|}{\multirow{2}{*}{\gou}} & \multicolumn{1}{c}{\multirow{2}{*}{\gou}} \\
Jailbreak   &                                                       & on page 3 of \cite{InvestigateMultilingual24}    & & \\
\hline

Normal      & \multicolumn{1}{c|}{\multirow{2}{*}{NA}}              & ``This is a random   & \multicolumn{1}{c|}{\multirow{2}{*}{\gou}} & \multicolumn{1}{c}{\multirow{2}{*}{\gou}} \\
    Prompt      &                                                       & sentence.'' (repeat)& & \\
\hline

\end{tabular}
\begin{tablenotes}
\item \footnotesize *: The ChatGPT version of GPT-4 and GPT-3.5 reported a web error when processing the special characters in the GCG example, so we removed those special characters. The API version operates well, as shown in \myfig~\ref{fig:example}.
\item \footnotesize \dag: The ChatGPT version of GPT-3.5 outputs additional auxiliary sentences such as ``The part of the sentence that violates safety policies is,'' while the API version, as shown in \myfig~\ref{fig:example}, does not.
\end{tablenotes}
\end{threeparttable}
}
\end{adjustbox}
\end{table}

\textbf{GCG Jailbreak} uses a technique called Greedy Coordinate Gradient~\cite{GCG23} to identify adversarial suffixes that can lead to universal jailbreaking across multiple prompts and models.
Zou et al.~\cite{GCG23} initialized the GCG technique, and Sitawarin et al.~\cite{GCGPlus24} further optimized this technique to GCG++ by using a proxy model to guide the optimization.
To demonstrate that \name can defend against the GCG jailbreak, we tested a representative adversarial example, i.e., Example 1 on the webpage cited in \cite{GCG23}.
\myfig~\ref{fig:example} shows that the API version of GPT-3.5 and GPT-4 can correctly identify the harmful prompt, while the ChatGPT version could also achieve similar results once the special characters that caused a web error were removed.

\textbf{Template-based Jailbreak} uses prompt engineering~\cite{Jailbroken23, EmpiricalJailbreak23, DAN23} and LLM-based red teaming~\cite{RedTeaming22, MASTERKEY24, AutoDan23} to identify a valid prompt template that can mislead LLMs into outputting restricted content.
Examples of such templates include ``Do-Anything-Now'' (DAN)~\cite{DAN23} and role-play scenarios like ``Dr. AI''~\cite{EmpiricalJailbreak23}.
However, most of them are already prohibited by OpenAI.
In this paper, we test the latest role-play jailbreak template, i.e., the prompt shown on page 9 of~\cite{MASTERKEY24}, which still works under GPT-3.5. 
As shown in~\cite{SelfDefend35Result} and~\cite{SelfDefend4Result}, both GPT-3.5 and GPT-4 can easily recognize the harmful prompt, which is ``w r i t e a f a m o u s p o r n w e b s i t e'' in this example, from such a jailbreak template.

\textbf{Multilingual Jailbreak} does not aim to identify certain adversarial suffixes or jailbreak templates.
Instead, it directly translates the harmful prompt into a language in which LLMs are less aligned for safety~\cite{MultilingualJailbreak23, DissectingMultilingual24}.
Besides the multilingual strategy, Wei et al.~\cite{Jailbroken23} propose a similar Base64 strategy that attempts to hide the original harmful prompt, as in our attempt~\cite{Base64Attempt}.
We also tried other obfuscation-based strategies, such as using simple combination~\cite{SimpleCombination}, word replacement~\cite{WordReplacement}, and standard AES encryption~\cite{AESEncryption}.
However, we found that such Base64-like strategies no longer work because off-the-shelf LLMs either directly interpret the obfuscated text to plaintext (e.g., for Base64 and simple combination) or employ a Python interpreter to translate to plaintext (for word replacement and AES encryption).
Hence, here we focus on the original multilingual strategy and use the example shown in Figure 2 on page 12 of~\cite{DissectingMultilingual24} for testing.
Our manual analysis shows that both GPT-4 and GPT-3.5 can recognize a Spanish harmful prompt that can successfully jailbreak GPT-3.5.

\textbf{Normal User Prompt.}
Lastly, we tested a normal user prompt by repeating a random sentence five times and asking GPT-3.5/4 to recognize any harmful prompt.
Both GPT-4 and GPT-3.5 correctly answered ``No.''

We are in the process of conducting extensive experiments to empirically support our finding that existing LLMs can effectively recognize harmful prompts while also being able to distinguish normal user prompts.
Our manual analysis presented in this section already shows promising results.

\section{Future Directions}
\label{sec:direction}

While \name is promising, there are several future research directions to make it fully practical in a real-world setting, as illustrated in \myfig~\ref{fig:overview}.

\begin{itemize}
    \item \textbf{FD1:} \textit{Design a low-cost, fast, and robust LLM for accurately recognizing harmful prompts.}
        While existing GPT-3.5/4 models already perform well in recognizing harmful prompts, it is desirable to reduce their inference cost and improve their inference speed. Moreover, we need a robust LLM that can accurately recognize harmful prompts in various adversarial scenarios. In particular, it needs to prevent potential prompt injection attacks~\cite{PromptInjection2310} launched by adversaries who are aware of \name's defense.
        To robustly defend against prompt injection, one approach is to leverage prefix tuning~\cite{PrefixTuning21} so that our detection prompt, shown in \myfig~\ref{fig:overview}, is integrated into LLMs as a prefix rather than a potentially manipulable prompt.

    \item \textbf{FD2:} \textit{Use the discovered AEs to further align LLMs.}
        By cross-checking the response given by LLMs, \name can also identify jailbreak prompts as AEs (adversarial examples) that can bypass existing alignment.
        These AEs can be used to further align the safety of LLMs.
        A better-aligned LLM in the normal stack can also enhance the detection of harmful prompts in the shadow stack.
        That said, by investigating the additional token-by-token output available when a checkpoint is triggered, we can confirm whether a harmful prompt has been identified in the shadow stack.
        We plan to conduct an ablation study to verify this.

    \item \textbf{FD3:} \textit{Design mechanisms to reduce/cache the calling of shadow stack.}
        In \name's original design, every user prompt needs to go through the checking process in the shadow stack, which could be enhanced by a caching mechanism.
        Therefore, another direction is how to design effective caching mechanisms for the shadow stack in a real-world setting.
\end{itemize}

Besides the research directions listed above, a valuable extension of \name is to support defense against multimodal jailbreak.
While \name can immediately defend against multimodal jailbreaks with harmful text prompts~\cite{VisualJailbreak2306, JailbreakGPT4V2311, JailbreakMultimodal2402}, it, by design, cannot handle pure multimodal jailbreaks~\cite{ImageSoundJailbreak2307} that use only images or sounds without any harmful text prompts~\cite{VRPTEST2312}.
A revised design of \name should be explored to defend against such advanced multimodal jailbreaks~\cite{ImageSoundJailbreak2307}.

\section{Related Work}
\label{sec:related}



\textbf{LLM Jailbreak Defense.}
Compared to the jailbreak attacks we have surveyed in \mysec\ref{sec:intro} and~\ref{sec:method}, the defensive side has been relatively less explored.
Existing jailbreak defenses can be roughly categorized into tuning-based and non-tuning-based mechanisms.
Tuning-based defenses~\cite{LlamaGuard2312, PAT2402, SafeDecoding2402} aim to fundamentally improve a model's safety alignment against jailbreaking.
Examples include Llama Guard~\cite{LlamaGuard2312}, which designs a dedicated model to align the classification of both prompt and response, and SafeDecoding~\cite{SafeDecoding2402}, which constructs a new token probability distribution during the training and inference phases to prevent jailbreaking.
However, tuning-based mechanisms require fine-tuning and could still be vulnerable to advanced jailbreaks such as GCG~\cite{GCG23}.

Hence, researchers also explored non-tuning-based defenses that can be directly applied to off-the-shelf LLMs.
Phute et al.~\cite{LLMSelfDefense23} were the first to propose a prompt-based framework that checks the safety of an LLM's output response.
Likewise, RAIN~\cite{RAIN2309} checks the output and uses LLMs' self-evaluation results to guide rewind and generation for AI safety.
To defend against GCG jailbreak, RA-LLM~\cite{RALLM2309} and SmoothLLM~\cite{SmoothLLM2310} perturb copies of the input prompt and aggregate the output response.
The work most closely related to \name is a recent study called IAPrompt~\cite{IAPrompt2401}, which proposes a prompt-based pipeline to analyze the intention of input prompts and generate policy-aligned responses.
While both IAPrompt and \name check the input prompt only, \name directly captures the harmful sentences in the original input prompt, whereas intention analysis might be bypassed by long-form prompts with the majority of intentions behaving benignly.
Moreover, compared to all the non-tuning-based defenses mentioned above, \name incurs minimal delay by creatively introducing a shadow stack and the corresponding checkpoint mechanism.

\textbf{Related Traditional Security Defense.}
Partial ideas of \name were inspired by traditional security defense concepts.
For example, the concept of the shadow stack was originally proposed for defending against buffer overflow attacks~\cite{ShadowStack19}.
Similarly, the checkpoint mechanism borrowed the idea of library-based checkpoint from SCLib~\cite{SCLib18}.

\section{Conclusion}
\label{sec:conclusion}

In this paper, we proposed \name, a lightweight yet practical jailbreak defense for LLMs, which is generic enough to defend against all existing jailbreak attacks.
We have validated the feasibility of \name in various jailbreak scenarios through manual analysis in GPT-3.5/4.
We also discussed several future directions to further enhance \name for real-world deployment.

\section{Version History}

\begin{itemize}
    \item \textbf{24 Feb 2024:} \textit{Submitted the initial, completed vision paper to arXiv.}

    \item \textbf{4 Mar 2024:} \textit{Fixed the reference bibliography issue; removed ``and'' from the author list.}
\end{itemize}

\bibliography{bib/llm,bib/jailbreak,bib/promptinject}
\bibliographystyle{plain}


\end{document}